\documentstyle[12pt]{article}

\topmargin - .5in \textwidth 6in
\begin{document}
\hspace{4.in} {\normalsize DOE/ER/40427-22-N96} \\
\begin{center}
     {\Large\bf Chiral Transparency}
\vskip 1.0cm
{\bf L. FRANKFURT}\\
{\it School of Physics and Astronomy, Tel Aviv Univ., 69978, Israel}\\
{\it Institute for Nuclear Physics, St. Petersburg, Russia }\\
{\bf T-S.H. LEE}\\
{\it Physics Div., Argonne National Lab, Argonne, IL 60439}\\
{\bf G. A. MILLER}\\
{\it Physics Department, BOX-351560, University of Washington}\\
{\it Seattle, Washington 98195-1560, USA}\\
{\bf M. STRIKMAN}\\
{\it Department of Physics,
Pennsylvania State University, University Park, PA 16802}\\
{\it Institute for Nuclear Physics, St. Petersburg, Russia} \\
\vskip .4cm
\end{center}
\vskip 1.5cm

\begin{abstract}
Color transparency is the vanishing of initial and final state interactions,
predicted by 
QCD to occur in high momentum transfer quasielastic nuclear reactions.
For specific reactions involving nucleons,
 the initial and final state interactions
are expected to be dominated by exchanges of pions. We argue that 
 these interactions are also suppressed in high 
momentum transfer nuclear quasielastic reactions;  this is 
``chiral transparency".
We show that 
that studies of the $e\;^3{\rm He}\to e'\;\Delta^{++}
nn$  reaction could reveal the influence of chiral transparency. 
\end{abstract}

\section{Introduction and Outline}

Effective Chiral Lagrangians have the same symmetries,
unitarity and cluster decomposition properties as QCD, and therefore
the two theories are expected to  yield the same predictions
\cite{W79}.  
Furthermore, a chiral perturbation
 theory treatment of the effective Lagrangians 
provides an organizing principle for
handling the very strong interactions typical of low momentum transfer
processes; see e.g. the reviews \cite{BKM95,picrev,rev3}). 

The utility of perturbation theory is due to a 
diminishing of the strong interaction which occurs        
as   a
straightforward consequence of approximate spontaneously broken
chiral symmetry.  
Suppose, we lived in a chiral world in which the 
up and down quark masses were exactly zero. In this world the 
pion would be a Goldstone boson and there are  theorems
\cite {theorem} that certain 
pion emission amplitudes
at threshold must vanish or be constant. 
 The pion mass is 
not actually zero---but it is small compared to typical hadronic 
scales.  This suggests that the amplitude can be described in a 
systematic manner as an expansion in ($k/\Lambda$), where
$k$ is a typical (small) momentum or energy scale $k\sim m_\pi$
and $\Lambda$ is a typical (large-mass) hadronic scale $\sim$ 1 GeV, 
such as $M_N$, $m_\rho$ or 4$\pi f_\pi$. This systematic expansion is 
called chiral perturbation theory ($\chi$PT).
This theory has been 
developed systematically for interactions of mesons 
and for interactions of mesons with a baryon; 
see the reviews\cite{BKM95}-\cite{rev3}. Furthermore, chiral perturbation
theory has been extended to systems 
of more than one  baryon \cite{W90,ORV94} so that now 
chiral perturbation theory can be used to describe 
the nucleon-nucleon strong force in 
a qualitative fashion\cite{ORV94}.

In principle, one ought to be able to apply the power counting in
a completely systematic
manner and work consistently to some given order.  However, it is
necessary
to introduce   counter terms to eliminate the divergent terms in
loop diagrams and therefore represent the short distance physics.
Once these counter terms are determined, one can predict
other observables.
 This approach has been used
with great success in describing the properties of the
pseudoscalar mesons\cite{GL84,GL85} as well as more recently baryons
\cite{JM91}.
Thus there seems to be a nice representation of low momentum transfer 
nuclear physics that is motivated by a fundamental theory.
 One uses pionic exchanges for the long range 
physics and the short distance physics is represented by 
counter terms.

The present 
 paper is concerned with a specific  extension of $\chi$PT to the regime 
of high momentum transfer physics in which a large momentum transfer
scale $Q^2$, much larger than the characteristic scale $\Lambda^2$ of 
$\chi$PT is introduced. 
Consider for example, elastic electron-proton scattering. At 
low momentum transfer one may 
try to describe the system in terms of the pion cloud of extent 
of the pion Compton radius ${1\over m_\pi}$. But when
$Q\gg m_\pi$ such effects do not enter. Instead one uses quarks 
and gluons to describe the interaction of the virtual photon with 
the proton. According to perturbative QCD, the high $Q^2$ process 
proceeds by components in which the quarks are close together.
Such components have been called point like configurations\cite{FMS85}.

The applicability of 
perturbative QCD to medium energy processes ($Q^2\sim 1- 5 GeV^2$)
has been questioned\cite{isguretc}. As a result 
three of us developed a criteria\cite{FMS92} to determine whether or not a 
point like configuration is formed in non-perturbative models of 
the nucleon. We found that point like configurations are formed for all 
realistic quark models - those 
in which there are correlations between the quarks. 
Point like configurations are also formed in the Skryme model, in 
which baryonic degrees of freedom are represented by pionic 
solitons. We especially recall that 
the nucleon's pion cloud provides negligible contributions to the
form factor 
if $Q\gg m_\pi$\cite{weise}.

The salient feature of 
point like 
configurations is that these do not 
 undergo  strong interactions for 
coherent low momentum transfer processes.
 This is because 
 small color (neutral) singlets have small forward scattering amplitudes.
As originally conceived within the two-gluon exchange model of the Pomeron
\cite{Low75}, this
results from the 
 sum of the gluon emission amplitudes cancelling
if the quarks and gluons of a 
color singlet are close together.  See Ref. \cite{FMS94}
 for further references and 
a discussion of how this cancellation is treated within 
QCD.

The new feature we wish to explore here is that   $\pi$ interactions
with a 
 color neutral point like configuration are
suppressed.  This is because 
the underlying interactions involved in producing or absorbing
a pion are also gluonic in origin. 
One example \cite{chanowitz81}
is the ratio of decay widths for the $\Upsilon'$ and
$\Psi'$ to decay to their ground states via pion emission. The ratios
are given by 
\begin{equation}
{\Gamma(\Upsilon'\to\Upsilon\pi\pi)\over \Gamma(\Psi'\to\Psi\pi\pi)}\approx
{<r^2_{\Upsilon'}>\over <r^2_{\Psi'}>}\approx {0.2^2\over 0.8^2}\approx 1/16.
\end{equation}
This ratio can be explained naturally using the idea that 
pion production arises from 
gluon emissions from the $b$ and $\bar b$ which tend to cancel.
Another very similar example occurs in the ratio 
\begin{equation}
{\Gamma(\Psi'\to\Psi\pi\pi)\over \Gamma(\rho'\to\rho\pi\pi)} \approx
135 KeV/200 MeV\cite{pdg96}.
\end{equation}
The small nature of this ratio was explained by Gottfried \cite{G78}
and Goldberg \cite{G75} using the long wavelength approximation to the
color multipole expansion  and the relative sizes of the charmed and
light quark systems. 
Another example, related 
to systems of only light quarks, comes from the work of 
deKam and Pirner \cite{pirner} in which the suppression of pion 
emission from bags of small size is used as a mechanism to provide stability 
against the collapse of the bag. Without this  suppression, the 
bag would collapse under the pressure     of pions outside the  bag. 
Weise et al \cite{weise2} invoked a similar mechanism by assuming the pion
quark coupling constant vanishes  at the center of the nucleon bag.
This was obtained by PCAC and the idea that the square of the pion mass
is proportional to the mass M of the quark, which depends on the distance (r) 
between the quark and the center of the nucleon. Motivated by the asymptotic
freedom idea that
light quarks are free and nearly massless when they are close
together, the function M(r)  was taken to vary as a
power of r. 
However, this freedom occurs only for color singlet systems
so that the motivation is very close to that for color 
transparency.
See Ref.~\cite{FMS85} for further discussion of the concept that pions are not
absorbed or emitted by point like configurations.

Thus
the notion that very small color neutral objects do not emit
pions seems to be consistent with diverse phenomena.
Consequently, point like configurations, produced  in high momentum transfer
quasielastic reactions, have no pionic cloud and are not expected to 
interact by pion exchange. We call this failure to interact 
``chiral transparency". For an early discussion of chiral transparency
see Ref.~\cite{FMS92}.

One possible example of chiral transparency occurs for the  process 
\begin{equation}
e+ {^3{\rm He}}\to\Delta^{++}(\vec q -\vec p_t) + n(\vec p_t) +n+e'.
\end{equation}
where $\vec q$ is the spatial momentum of the incident virtual photon, 
$\vec p_t$ is the transverse $(\vec p_t\cdot\hat q = 0)$ momentum of the
detected neutron.  Under chiral transparency the cross section
vanishes at large values of momentum transfer.  Another example, is
that the production of a non-resonant uncorrelated proton and a
$\pi^+$
is also expected to be suppressed.

To see how this suppression may come about, start by first considering
the 
conventional approach. One expects that this reaction proceeds
by various terms.  The virtual photon could land on the proton ($p$)
converting it to a high momentum $p$ or $\Delta^+$.  The $p$ or $\Delta^+$
then undergoes a charge exchange reaction, i.e. $pp\to\Delta^{++}n$
or $\Delta^+p\to\Delta^{++}n$.  Such reactions are dominated by pion
exchanges\cite{PICEX}.
  The photon could also be absorbed by a contact interaction
of the form $\gamma^*p\to\Delta^{++}\pi^-$.  In both of these processes
the reaction proceeds by $\pi$ exchange in the final state.  Another
process would involve a $\Delta^{++}N$ component of the initial
wave function.  
  Such components are strongly suppressed by the $\Delta$
mass and systematic searches for such components have never
succeeded. 
The relatively large mass of the 
$\Delta$, combined with the effects of short range repulsions,
causes a vast reduction in the influence of $\Delta's$ in the 
initial state as compared with those  produced by final state 
interactions.

Under chiral transparency the absorption of the virtual photon forms a
point like configuration, PLC,
 which cannot emit a pion.  In that case the cross section for
 quasielastic production of the  $\Delta^{++}$ would vanish. This is 
chiral transparency.  There is a complication because the point like
configuration 
expands as it moves, so that it may indeed 
emit  a pion some distance away from the point where it is produced.
This physics is modeled by allowing the 
$\pi$-coupling constant 
to be a function of the
propagation length, 
see below.

At Q$^2$ high enough to produce a PLC $Q^2\sim 1\; GeV^2$
, but not very large,   the momentum  
of the 
produced PLC is small and the PLC expands to full
baryonic size.  The $\pi$-coupling is therefore of normal strength.
However, as Q$^2$ increases, Lorentz time dilation takes over 
and the $\pi$-coupling is
reduced and the $\Delta^{++}$ production cross section is suppressed.
Thus the
basic idea is that under chiral transparency the cross section goes 
to zero, as 
Q$^2$ is increased, much faster than does a conventionally computed cross
section.

The production of a $\Delta^{++}$  can also proceed via $\rho$ meson 
 exchange.
But the $\rho$ coupling to the  point like configuration is also
expected to 
be suppressed. Thus although 
 the relative importance of  $\pi$ and $\rho$  meson exchange
is somewhat model dependent, transparency effects should occur
for  either 
either  meson exchange. Our
numerical work is based on the detailed baryon-baryon interaction 
model of Lee\cite{LEE83}. In 
that model, the pion exchange effects are much more
important than are those of rho meson exchange. Thus we shall ignore
the
effects of $\rho$ meson exchange in the remainder of the present paper.

The net result of this is that  if the high  momentum transfer
virtual  photon interacts 
with a proton bound in a nucleus a point like configuration is 
produced which does not interact with the surrounding 
nucleons. Thus the chiral physics undergoes a qualitative 
change; the interaction vertex is suppressed  by a factor smaller 
than the usual  $k/\Lambda$. 
The crucial issue is the
 value of $Q^2$ needed to turn on this transparency. 
If the value 
$Q^2$ were too high, chiral transparency would have no observable effect.
On the other hand, if this number were too
low, chiral perturbation theory would have diminished
relevance because 
quark physics would enter at lower momenta than expected.

There are at least two major practical 
problems in observing chiral transparency.
The first is that the PLC must not expand too fast.
This disruptive effect  can be reduced 
by using the smallest target nucleus possible-this 
is $^3He$. An additional advantage of using this target is that 
good wave functions for the ground state are available.

The second problem is that one needs to use the a reasonably accurate 
version of 
the conventional process of $\Delta^{++}$ production. 
%It is necessary to start by having a good representation of the conventional
%cross section.  
Lee's\cite{LEE83} 
interaction  accounts for the real and imaginary parts of 
nucleon-nucleon phase shifts,
and mixing parameters for  energies up
to $s_{NN} = 8\;\rm GeV^2$
 This is high enough for the present purpose of providing motivation for an
experiment at the Jefferson Laboratory 
because
$Q^2=(s_{NN}-4 M_N^2)/(1 +2{M_N-M_\Delta\over M_\Delta})$
for quasielastic production of a $\Delta$.

The basic idea is that at Q$^2$ near 1 GeV$^2$  we expect that the 
conventional theory will work. This can be tested by  comparing with 
relevant data when such becomes available. As Q$^2$ increases from 1
to $\sim$ 6 GeV$^2$ one expects that
chiral transparency will become important. We shall
 compute the ratio of cross sections with and
without the effects of chiral transparency.

It is necessary to consider the question of whether or not the
existence of PLC  is already ruled out. If so, chiral transparency 
could not exist. 
Thus, it is worthwhile to briefly review the current status of the color 
transparency experiments most closely related to the  electron-nuclear
interaction of the present paper. Color transparency (CT) and color
coherent effects have been recently under intense experimental and
theoretical investigation.  The (p,pp) experiment of Carroll et al.
\cite{C88} found evidence for color transparency  \cite{JM93} while
the NE18 (e,e'p) experiment \cite{Makins94} did not.  The appearance
of color transparency depends on formation of a point-like
configuration (PLC)  by hard scattering.   The Q$^2$ of the NE18
experiment $(1\leq Q^2\leq 7$ GeV$^2$) seem to be large enough to form
a small color singlet object.   We believe that this failure to observe 
significant color transparency effects is  caused by
the rapid expansion of the  point like configuration  to nearly normal
size (and  nearly normal absorption ) at the relatively low  momenta
of the ejected protons \cite{F88},\cite{JM90}. In particular, models of color
transparency which reproduce the (p,2p) data and include expansion
effects predicted small CT effects for the NE18 kinematics, consistent
with their findings, see the  discussion in Ref.\cite{FMS94}.

Thus the ability to  observe color transparency effects  at
intermediate values of Q$^2$ (between 1 and 7 GeV$^2$ ) rests on
finding ways to avoid the effects of PLC expansion.  One may use 
light nuclei and kinematics that require double scattering  Ref.
\cite{E94,FMS95}. The idea is that color transparency effects suppress the
double scattering terms so that  the cross section vanishes, in 
contrast with the  predictions of the usual  Glauber model.  

The ability to observe chiral transparency rests also on 
requiring a final state interaction mediated by pion exchange,
 in this case a charge exchange 
reaction. We immediately take the target nucleus to be small, 
so that the expansion of the PLC 
should not play a dominant role. Thus the observation of a significant cross
section at  large values of $Q^2$ would rule out the existence of 
chiral transparency.

We outline the remainder of the paper. Section 2 is concerned 
with a formal but schematic derivation of chiral transparency. Section 3
deals with the 
development of formulae necessary to compute cross 
sections both in the conventional approach and using the chiral 
transparency idea. The results are presented in Section 4, and 
summarized and discussed in Section 5.

\section{Chiral Transparency}

This section is intended to further specify the assumptions that underlie
chiral transparency. We shall use a schematic notation to simplify the 
discussion as much as possible. Our starting point is the description of the
nucleon wave function in the hadronic  Fock state basis:
\begin{equation}
|N>=\sqrt{Z}\left[|N>_0+C_{N,\pi}|N,\pi>_0+C_{\Delta,\pi}|\Delta,\pi>_0+\cdots
\right]\label{state}
\end{equation}
The states labelled with subscript 0 are eigenstates of the Hamiltonian $H_0$
given by
\begin{equation}
H_0=H-H_{\pi,q}-\tilde V,
\end{equation}
where H is the complete Hamiltonian, $H_{\pi,q}$ represents the pion  quark
interaction, and $\tilde V$ is that part of the two (or three ) nucleon force
that is not generated by various iterations of $H_{\pi,q}$.
The interaction $H_{\pi,q}$ can be taken to be pseudovector
pion-quark coupling, with the modification that there is no interaction
when all quarks in the system are at the same location. For example, 
\begin{equation}
H_{\pi,q}={1\over 2 f_\pi}
\int\;d^3r {r^2\over <r^2>}
\bar \psi(\vec r)\gamma_\mu\gamma_5 \tau\cdot \partial^\mu
\phi_\pi(\vec r)\psi(\vec r). \label{ill}
\end{equation}
Here $\psi$ and $\phi_\pi(r)$ are the quark and pion field operators, 
the center of the nucleon is at $\vec r=0$,
and 
$<r^2>$ is the nucleonic expectation value of the $r^2$. 
%%answer to lonya
The suppression factor $r^2/<r^2>$ arises by starting with
an operator $\propto\sum_{i\ne j}^3 (\vec r_i-\vec r_j)^2$ and assuming
spherical wave functions used in a mean field approximation so that the
$\vec r_i\cdot\vec r_j$ term averages to zero. 
The particular form $r^2/<r^2>$
is meant to be a schematic representation of any function of $r^2$ and $<r^2>$
that vanishes at $\vec r=0$
and has an expectation value of unity in the nucleon wave
function. Equation (\ref{ill}) is meant only as an illustration; a complete
derivation of the pion-quark interaction is beyond the scope of this paper.

The ability of the Fock state expansion of eq(\ref{state}) to represent 
the nucleon wave function in a few terms (in particular, ignoring states with 
two or more pions) depends on underlying dynamical assumptions. In this paper
we use Lee's 
model\cite{LEE83} in which the pion-nucleon form factor is a dipole with 
from the cloudy bag model because a dipole form factor 
with $\Lambda$ =650 MeV. This soft form factor 
corresponds to a large three-quark confinement (bag) radius of 1.33 fm.
In this case, cloudy bag model studies \cite{cbm} show that it very safe
to ignore states with two  or more pions and that $Z\approx 1$.

We now turn to how the nucleon state of eq.(~\ref{state}) 
responds to an external
electromagnetic probe denoted by $T_H(Q^2)$. Numerical work \cite{weise}
indicates that the contribution of the pion cloud is negligible 
for values $Q^2$ greater
than about 0.5 (GeV/c)$^2$. Thus we may write 
\begin{equation}
T_H(Q^2)|N>\approx \sqrt{Z}|N>_0,\;\;\qquad Q^2> 0.5\; (GeV/c)^2.\label{pigone}
\end{equation}
We note that
$_0<N|H_{\pi,q}|N>_0 $ gives the ordinary (lowest order) 
 pion-nucleon emission  vertex, which is not suppressed. 

Chiral
 transparency requires  higher values of $Q^2$. We rely on earlier work
\cite{FMS92} which indicates that 
\begin{equation}
T_H(Q^2)|N>\approx |PLC>,\;\; \qquad Q^2 > 1\; (GeV/c)^2, \label{PLC}
\end{equation}
where $|PLC>$ represents a point like configuration, one in which the quarks
are close enough together for significant suppression of the pion quark
interaction to occur. In particular, the pion cloud absent according to 
Eq.~(\ref{pigone}) remains absent at higher values of $Q^2$. In particular,
Eq.~(\ref{PLC}) implies that
\begin{equation}
H_{\pi,q}|PLC>=0. \label{chiralt}
\end{equation}
This is the formal statement of chiral transparency.

The state $|PLC>$  is not an eigenstate of the Hamiltonian, so it will
evolve to another state, one which is necessarily larger and which therefore
interacts via $H_{\pi,q}$. Thus we need to consider the time evolution.
Suppose a PLC is produced at a position $\vec r$ and moves to  a position 
$\vec r\;'$. This requires a time $\tau\approx |\vec r\;'-\vec r|$,
since we are interested in rapidly moving PLC's. The effects of this time 
evolution can be incorporated in by using the Heisenberg representation so that
the time-dependent pion-quark interaction 
$H_{\pi,q}(|\vec r\;'-\vec r|)$ is given by
\begin{equation}
H_{\pi,q}(|\vec r\;'-\vec r|)=e^{iH_0|\vec r\;'-\vec r|)} H_{\pi,q}
e^{-iH_0|\vec r\;'-\vec r|)} . \label{evolve}
\end{equation} 
The relevant matrix elements for producing or absorbing pions is then 
$$_0<B|H_{\pi,q}(|\vec r\;'-\vec r|)|B'>_0,$$
 where B,B' represents the 
baryonic states $N,\Delta\cdots$. This quantity is determined 
by a coupling constant which depends on $|\vec r\;'-\vec r|$:
$g_{\pi\;B,B'}(|\vec r\;'-\vec r|)$.
The explicit evaluation of Eq.~(\ref{evolve}) must necessarily involve many
detailed model assumptions, and is not a subject of the present work. Instead
we rely on the related experience of color transparency in which the
the explicit evaluation of an equation similar to Eq.~(\ref{evolve})
using a 
sufficiently large hadronic basis \cite{JM92} yielded results similar
to that of a model based on quantum diffusion (qdm)\cite{F88}.
The notion behind the  qdm is that the interaction of the produced PLC  is 
proportional to $|\vec r\;'-\vec r|$ for $\vec r\;'\approx\vec r$, but
approaches normal strength  for larger
values of $|\vec r\;'-\vec r|$. Furthermore, 
the size (and interaction strength) of the initially produced PLC depends on
$Q^2$.
With these features in mind we assume a form:
\begin{equation}
g_{\pi\;B,B'}(|\vec r\;'-\vec r|)=g^0_{\pi\;B,B'}
\left(1-\kappa e^{-|\vec r\;'-\vec r|/l_c}\right), \label{gr}
\end{equation}
where $l_c$ is the length (or time) scale required for
the PLC to evolve to a configuration of nearly normal hadronic size.
The idea that this length is given by  the Lorentz time dilation of
a relevant rest frame time  leads to
\begin{equation}
l_c = {2\;p\over \mu^2},
\end{equation}
where $p$ is
the longitudinal momentum of the PLC ($\approx |\vec q\;|)$.
 In the (e,e'p) reaction
 $\mu^2\approx$ 0.7 GeV$^2$, but 
in the present situation $\mu$ may  be smaller than that. We estimate 
that $l_c\approx r_\pi {p\over m_N}$ which arises from using the 
pion radius 
as the rest frame time for expansion and ${p\over m_N}$ as the time dilation 
factor. This gives $\mu^2=m_n/r_\pi\approx 0.3\; GeV^2$.
The parameter $\kappa$ is related to the feature that high $Q^2$ is
required to form a PLC that does not emit pions. We use the form
\begin{eqnarray}
\lambda_0&=& 1 \nonumber \\
\kappa &=& 0 \, \, , \qquad Q^2 < Q_0^2 \nonumber \\
       &=& 1 - \frac{Q_0^2}{Q^2} \, \, , \qquad Q^2 \geq Q_0^2 \, , \nonumber
\end{eqnarray}
$Q_0^2$ is a parameter that controls the momentum transfer at which 
the PLC is assumed to be formed. In particular, if $Q_2<Q^2_0$ the
pion-baryon coupling constants have their normal strength and chiral 
transparency does not occur. We shall use several values of $Q_0^2$ in
this initial investigation.

We need to consider how to use Eq.~(\ref{gr})  in computing the relevant
scattering amplitudes. Consider first the amplitude ${\cal M}$ 
for the ordinary process 
in which the $\Delta^{++}$ is produced by the absorption of a photon 
followed by a final state charge exchange operator. Then 
schematically 
\begin{equation}
{\cal M}\sim\sum_{B'}\int d^3r_1'd^3r_1d^3r_2
\Psi_f^*(\vec r_1\;',\vec r_2)  V_{\Delta^{++}n;B'^+,p}(|\vec r_1\;'-\vec
r_2|)G_{B'}(|\vec r_1\;'-\vec r_1|)<B'|T_H(Q^2)|p> \Psi_i(\vec r_1,\vec
r_2),\label{m1}
\end{equation}
where $\Psi_i$ represents the initial wave function of the bound pp
system and  $\Psi_f$ represents the final state $\Delta^{++}n$ wave
function. 
The final state charge exchange interaction, $V_{\Delta^{++}n;B'^+,p}$
contains the effects of the pion-baryon operator generated by the 
pion quark interaction Hamiltonian. The input wave functions and
charge exchange interaction are specified in the next section.

Under chiral transparency one
uses Eq.~(\ref{evolve}) and the related ansatz Eq.~(\ref{gr}). In the quantum
diffusion model the
argument of $g^0_{\pi\;B,B'}$ is the same as that of the intermediate 
baryonic Green's function, $G_B$, so that 
the effects of chiral transparency can be included simply by
multiplying $G_B $ by 
$g^0_{\pi\;B,B'}(|\vec r\;'-\vec r|)/g^0_{\pi\;B,B'}(0).$   In
operational terms one uses a quantity 
\begin{equation}
G^\chi_{B'}(|\vec r\;'-\vec r|)=G_{B'}(|\vec r\;'-\vec
r|){g^0_{\pi\;B,B'}(|\vec r\;'-\vec r|)\over g^0_{\pi\;B,B'}(0)}.\label{gchi}
\end{equation}
Thus when
 chiral transparency is invoked, the amplitude is ${\cal M}^{\chi}$ given by
\begin{equation}
{\cal M}^{\chi}\sim\sum_{B'}\int d^3r_1'd^3r_1d^3r_2
\Psi_f^*(\vec r_1\;',\vec r_2)  V_{\Delta^{++}n;B'^+,p}(|\vec r_1'-\vec
r_2|)G_{B'}^\chi(|\vec r_1'-\vec r_1|)<B'|T_H(Q^2)|p> \Psi_i(\vec r_1,\vec
r_2).\label{mchi}
\end{equation}

The difference between Eqs. (\ref{m1}) and (\ref{mchi}) is that in the latter
equation the 
pion-baryon interaction is suppressed for small values of $|\vec r\;'-\vec r|$.
This effect is carried by the different Green's functions,  in
the present simple version of the theory.

\section{Detailed Formalism}

So far we have been concerned with the a general description of chiral
transparency. To make further progress, it is necessary to choose a specific
reaction and display how the numerical results are obtained.
The absorption of the photon by a $^1S_0$ $pp$-pair in
$^3He$ is the example we choose because the target is the 
lightest stable one with
two bound protons.

We proceed by first describing the formalism for the conventional treatment.
The effects of chiral transparency are included simply by replacing
the propagator of the intermediate baryon $G_{B'}$ by 
$G^\chi_{B'}$ according to Eqs.~(\ref{gchi}) and (\ref{mchi}).

The absorption of the photon by a $^1S_0$ $pp$-pair in
$^3He$ is governed by a 
basic mechanism:  $ \gamma+(pp) \rightarrow B p \rightarrow
\Delta^{++} n$ with
$B=p, \Delta^+$. The amplitude for this transition
\begin{eqnarray}
T = \sum_{B=p, \Delta^+} <\phi_{pp} | J_{\gamma p, B}\cdot
\epsilon_{\lambda} | B p >
G_{Bp}(E) < B p | t(E) |\Delta^{++} n > \, , \label{dynamics}
\end{eqnarray}
where $\epsilon_{\lambda}$ is the photon polarization vector,
$\phi_{pp}$ the wavefunction of the
$pp$-pair, $J_{\gamma p, B}$ a one-body transition current operator,
and $t$ the two-baryon transition t-matrix.  (Our notation  here is that
$T={\cal M}^*$.)

The current $J_{\gamma p, p}$ is the standard empirical proton
form, while $J_{\gamma p , \Delta^+}$ was determined in studies of
$N(e,e'\pi)$ reaction Refs.~\cite{NL91} and \cite{Lee94}
% Nozawa and Lee, 1991, and Lee, 1994).
 The 
$B^\prime N \rightarrow \Delta N$ transition amplitude  ( a 3 by 3 matrix)
$t(E)$ is calculated from a 
unitary model\cite{LEE83}
 of the coupled $NN\oplus N\Delta \oplus \pi NN$ reactions.

The produced $\Delta^{++}$ decays  into a detected $\pi^+ p$ state.
Thus 
the total production cross section
must be integrated over all of the possible invariant 
masses of the resonant $\pi^+ p$ system.  
The invariant mass $W$ of the final coupled $\Delta N \oplus \pi NN$ system
is related to the
photon four momentum $q=(\omega,\vec{q})$ in $^3He$ rest frame and 
$q_c=(\omega_c,\vec{q}_c)$ in the $\gamma^*-BB$ c.m. frame by
\begin{equation}
W = \omega_c + E_d\,(\vec{q}_c ) = (\omega + M_d )^2 - \vec{q}\,^2
\end{equation}
The photon 
four-momentum satisfies the relation $ q^2  =\omega^2 - \vec{q}^2
\equiv q_c^2 = \omega^2_c -\vec{q}^{2}_{c}$. Thus the mass of the 
final state $\Delta^{++}(\to p, \pi^+)$ can vary between the ranges given by 
$W-m_n \geq m_\Delta \geq m_\pi + m_n$, so that the total production 
cross section is given by
\begin{equation}
{{d\sigma \over d\Omega}}_{c.m.} = \int^{W-m_n}_{m_n + m_{\pi}} \,dm_{\Delta}\,
{d\sigma \over d\Omega\,dm_{\Delta}}\,(W)
\end{equation}

The cross section for 
production of a $\Delta^{++}n$ final state  from an initial pp pair
with a relative wavefunction 
$\phi^{T,M_T}_{[LS]J,M_J}$  can be derived in a straightforward way
from 
Eq.~(\ref{dynamics}). The result is 
\begin{eqnarray}
 {d\sigma \over d\Omega\,dm_{\Delta}} (W)&& = {(2\pi )^4 \over v_i}\,
{1 \over (2J + 1) \cdot 2} \sum_{m}
\rho_{N\Delta} (p_0) \rho_{\pi N} (p_0,k_0)\nonumber \\
&&\times | \langle\,\vec{q}\,\lambda\,,JM\,,TM_T\, | T |\, 
\vec{p}_0\, m_{s_{\Delta}}
m_{\tau_{\Delta}} m_{s_2} m_{\tau_2}\,\rangle\,|^2\label{cross}
\end{eqnarray}
where $v_i = (q_c/\omega_c + q_c/E_d(q_c))$ is the relative velocity 
between the photon and the initial $NN$ pair. The $N\Delta$ relative momentum
$\vec{p}_0$ and the $\pi N$ relative momentum $\vec{k}_0$ are defined by their
corresponding invariant masses $W$ and $m_\Delta$ 
\begin{eqnarray}
p_0 &=& {1 \over 2W} 
\left[ (W^2 - m^{2}_{\Delta} - m^{2}_{N} )^2 - 4\, m^{2}_{\Delta}\,
m^{2}_{N} \right] ^{1/2}
\nonumber \\
k_0 &=& {1\over 2m_{\Delta}} \left[ (m_{\Delta} - m^{2}_{N} - m^{2}_{\pi} )^2 -
4\, m^{2}_{N}\,m^{2}_{\pi} \right] ^{1/2} 
\end{eqnarray}
In Eq.(\ref{cross}), we have introduced the phase-space factors for the final
$N\Delta \oplus \pi NN$ subsystem. By using a dynamical description
of the $N\Delta$ propagator and the $\Delta\rightarrow \pi N$ vertex determined
in Ref.\cite{LEE83}, we obtain
\begin{equation}
\rho_{N\Delta} (p_0) 
= {E_N (p_0)\, E_{\Delta} (p_0)\,p_0 \over E_N (p_0) + E_{\Delta}(p_0)}\,
\end{equation}
and
\begin{equation}
\rho_{\pi N} (p_0,k_0) 
= {E_{\pi} (k_0) E_N (k_0) k_0\over E_N (k_0) + E_{\pi} (k_0)} \,|\,
{h(k_0) \over W-E_N (p_0) - E_{\Delta} (p_0) - \Sigma_{\Delta} (p_0,W)}\, |^2
\end{equation}
where $\Sigma(p_0,W)$ is the complex $\Delta$ self energy evaluated 
in the presence of a spectator 
nucleon, $h(k_0)$ is a $\Delta\rightarrow \pi N$ vertex function. This
is a dipole with $\Lambda=650 $ MeV.
The heart of the calculation is the $\gamma^*pp\rightarrow  Bp \rightarrow 
\Delta n$ transition amplitude. Explicitly, we have 
\begin{eqnarray}
&&\langle\, \vec{q}\, \lambda\,, JM\, TM_T  |T(W)|\, 
m_{s_{\Delta}} m_{\tau_{\Delta}} m_{s_2} m_{\tau _2} \vec{p}\,\,\rangle 
\nonumber \\
&& = \sum_{B= p,\Delta^+}\sum_{m} \langle\,JM| L'S'M^{'}_{L} M^{'}_{s}\,
\rangle\, \langle\, S' M^{'}_{s}
|s_1^\prime,s_2^\prime, m^{'}_{s_1} \,m^{'}_{s_2} \,\rangle
\langle\, TM_T\,|\,\tau_1^\prime\tau_2^\prime,\,1/2\,m^{'}_{\tau_1} 
m^{'}_{\tau_2} \,\rangle
\nonumber \\
&& \times \int\, d\vec{p}\,' \phi^{J}_{L'S'}\left(\,|\vec{p}\,' - {\vec{q} 
\over 2}\,| \right)
 Y_{L'M'_L} \left( \widehat{p' - {q \over 2}} \right)
 \langle\,\vec{p}\,' - \vec{q}\,, m^{'}_{s_1} m^{'}_{\tau_1}\,|\, 
J_{\gamma N, B}\,| \vec{p}\,' m^{'}_{s_{B}} m^{'}_{\tau_{B}}\,\rangle
\nonumber \\
&& \times G_{Bp} (\vec{p}\,,W)
\langle\,\vec{p}\,' m^{'}_{s_{B}} m^{'}_{\tau_{B}} m^{'}_{s_2} m^{'}_{\tau_2}
\,| t(W)\,| \vec{p}_0\,m_{s_{\Delta}} m_{\tau_{\Delta}} m_{s_2} m_{\tau_2}\,
\rangle
\label{big}\end{eqnarray}
In the above equation, $\phi^J_{LS}(p)$ is the radial part of the initial
$pp$ relative wavefunction. 

We now discuss the propagators appearing in Eqs.(\ref{dynamics}) and
(\ref{big}). 
The $N\Delta$ propagator is defined by 
\begin{equation}
G_{N\Delta} (\vec{p}\,'\,,W) = 
{1 \over W-E_N (p') - E_{\Delta} (p') - \Sigma_{\Delta}
(\vec{p}\,'\,,W)},\label{gd}
\end{equation}
and the intermediate NN propagator by 
\begin{equation}
G_{NN} (\vec{p}\,'\,,W) = {1 \over W-E_N (p') - E_N (p') +i\epsilon}.\label{gn}
\end{equation}

This completes the specification of the conventional theory, so that
we may 
now turn to the effects of chiral transparency. As discussed above, we may
 simply 
change the coupling constant so that its growth with the expansion
of the wave packet is modeled as a function of the propagation length.
The present formalism is in momentum space so that we 
need to  re-express Eq.(\ref{gchi}) in momentum space.
To do this we first note that Eqs.(\ref{gd}),(\ref{gn})
can be re-expressed as 
\begin{equation}
G_{N,\alpha}(\vec {p}\,'\,W)=
{C(p_0^2(\alpha))\over p_0^2(\alpha)-p'^2+i\epsilon},\;
\alpha=N,\Delta, \label  {G}
\qquad\end{equation}
where 
$p_0(N,\Delta)$ is the position of the pole of $p'$. The 
Fourier transform of these Green's functions $\sim e^{ip_0r}/r$. Under
chiral transparency these are replaced by 
$\sim e^{ip_0r}/r(1-\kappa e^{-r/l_c})$.
This means that the effects of chiral transparency can be included by 
replacing the above $G_{N,\alpha}$ by $G_{N,\alpha}^\chi$ given by 
\begin{equation}
G_{N,\alpha}^\chi(\vec {p}\,'\,W)=
{C(p_0^2(\alpha))\over p_0^2-p'^2+i\epsilon}
-{\kappa 
C(p_0^2(\alpha))\over (p_0+i/l_c)^2(\alpha)-p'^2+i\epsilon}.
\label{gpchi}
\end{equation}
%%answer to lonya, yes you are right
%%formula changed- not to worry harry did not use this
The transition matrix can be calculated according to Eq. (\ref{G}) to obtain 
results for the conventional theory, 
or by using Eq.(\ref{gpchi}) to obtain the results for chiral transparency.

\section{Results}
We now discuss the results of explicit calculations which are focussed on
the quasielastic production of the $\Delta^{++}n$ system. That is 
both initial protons are taken to have momentum nearly equal to zero,
i.e. at the peak of the initial state wave function, so that the cross
section is maximized.
This is achieved by setting 
the invariant mass of
the produced $\Delta^{++}$ equal to the physical
$\Delta$ mass of 
$1236$ MeV, and 
by setting the momentum of the produced neutron $p$ according
to the relativistic constraint
\begin{eqnarray}
\alpha = \frac{\sqrt{m_N^2 + p^2} - p_z}{m_N} =1
\end{eqnarray}
where $p_z$ is the neutron momentum projected in the direction of photon.
Standard electron scattering
kinematics are used. 
Setting  the energy-transfer $\omega \sim 300$ MeV of the virtual photon
to emphasize the excitation of a proton to a $\Delta$ state, choosing $Q^2$ and
using Eq. (28) to determine the angle between the outgoing $\Delta$ and the
incident virtual photon determines the necessary kinematics.
We shall present our results in the form of a ratio of cross sections
defined by 
\begin{equation}
CT\equiv {d\sigma^\chi\over d\sigma} .\label{tchi}
\end{equation}.

The first set of results,  shown in Fig.~1, are determined by the 
choice $Q_0^2=0.55$ GeV$^2$, and the variation with $\mu^2$ is
displayed. Harmonic oscillator wave functions are
used with b=1 fm. The ratio $CT$ differs significantly from zero
at fairly low values of $Q^2$ for all of the values of $\mu^2$
displayed.  

The dependence on $Q_0^2$ is displayed in Fig.~2. 
This parameter determines the $Q^2$  necessary for a PLC to
be formed and for chiral
transparency to set in. We have argued above and elsewhere that
Q$^2_0\approx 1$ GeV$^2$, but the results shown in Fig.~2 indicate
that experiments could determine whether or not the 1 GeV$^2$ estimate
is anywhere near correct.

The sensitivity to the initial pp wave function can be studied by
examining 
Fig. (3).  This figure shows the sensitivity to the
harmonic oscillator parameter b. Increasing this number increases the size of
the system, and therefore also the 
influence of PLC expansion. One therefore expects that the value of 
$CT$ increases with increasing b. This is indeed the case.
The figure also compares the effects of using harmonic
oscillator wave functions with those of using the Paris s-state deuteron wave
function.   This is done to examine the influence of  short range correlations.
While there is sensitivity to using different wave functions,
the significant feature is that 
$CT$ can be much less than unity for all of the wave functions studied.

The net result is that the predicted effects of chiral transparency
seem to be  very significant.

\section{Summary}

Color transparency involves the suppression of initial or final state 
interactions.  This effect enhances (e,e'p) and (p,pp) reactions but
suppresses reactions (e,e'pp) and (e,e'$\Delta^{++}$ n) in which a second
interaction is required for the reaction to occur.  The production of a
$\Delta^{++}$ requires a $\pi^-$ emission.  The suppression of this
process represents an example of chiral transparency.  Measuring chiral
transparency effects would provide evidence that high Q$^2$ 
electron-proton scattering 
proceeds via the formation of a PLC.  Potentially large signals are
available at Jefferson Lab.

This work is partially supported by the U.S. Department of Energy, Nuclear
Physics Division
(grants DE-FG06-88ER 40427, DE-FG02-93-ER 40771,
 and contract  W-31-109-ENG-38) and by the
US-Israeli Bi-National Science Foundation grant -9200126.

\begin{figure}
\caption{Chiral transparency ratio $CT$ of Eq. (\ref{tchi}).
The transverse momentum of the neutron is 0.3 GeV/c. Harmonic
oscillator wave functions are used with b=1 fm. The parameter
$\mu^2$, which determines the value of $l_c$, is varied.}
\end{figure}
\begin{figure}
\caption{Chiral transparency ratio $CT$ of Eq. (\ref{tchi}).
The transverse momentum of the neutron is 0.3 GeV/c. Harmonic
oscillator wave functions are used with b=1 fm. The parameter
$Q^2_0$, which determines the value of $\kappa$, of eq.~(\ref{gr}) is
varied.}
\end{figure}

\begin{figure}
\caption{Chiral transparency ratio $CT$ of Eq. (\ref{tchi}).
The transverse momentum of the neutron is 0.3 GeV/c. Harmonic
oscillator wave functions are used and the variation with b
is studied. Results obtained by using 
the Paris s-state deuteron wave
function are shown.}
\end{figure}

\end{document}